\documentclass[11pt,preprint]{aastex}

\newcommand{\masr}{mas~yr$^{-1}$}

\slugcomment{to appear ApJ, 597, Nov 10, 2003}

\shortauthors{Hardee}
\shorttitle{Modeling Jet Structures}

\begin{document}

\input colordvi.sty
 
\baselineskip 12pt
\parskip 0pt


\title{Modeling Helical Structures in Relativistic Jets}

\author{Philip E. Hardee}
\affil{Department of Physics \& Astronomy, The University of Alabama,
Tuscaloosa, AL 35487}
\email{hardee@athena.astr.ua.edu}



\begin{abstract}
\baselineskip 12pt

Many jets exhibit twisted helical structures.  Where superluminal
motions are detected, jet orientation and pattern/flow speed are
considerably constrained.  In this case modeling efforts can place
strong limits on conditions in the jet and in the external
environment.  This can be done by modeling the spatial development of
helical structures which are sensitively dependent on these
conditions.  Along an expanding jet this sensitivity manifests itself
in predictable changes in pattern speed and observed wavelength.   In
general, twists of low frequency relative to the local resonant
frequency are advected along the expanding jet into a region in which
the twist frequency is high relative to the local resonant frequency.
The wave speed can be very different in these two frequency regimes.
Potential effects include helical twists with a nearly constant
apparent wavelength, an apparent wavelength scaling approximately with
the jet radius for up to two orders of magnitude of jet expansion, or
multiple twist wavelengths with vastly different intrinsic scale and
vastly different wave speeds that give rise to similar observed twist
wavelengths but with very different observed motion.  In this paper I
illustrate the basic intrinsic and observed behavior of these
structures and show how to place constraints on jet conditions in
superluminal jets using the apparent structures and motions in the
inner 3C\,120 jet.

\end{abstract}

\keywords{galaxies: jets --- hydrodynamics --- relativity --- galaxies: active --- galaxies: individual (3C\,120)
\vspace{-0.8cm}
}

\baselineskip 12pt
\parskip 2pt

\section{Introduction}

The resolved highly collimated relativistic AGN jets can exhibit
twisted time-dependent structures. For example, the 3C\,345 jet shows
non-radial curved motions and accelerations that have been interpreted
as motions along a helix lying on the surface of a cone
\citep{ZCU95,S95}, the jet in M\,87 shows helically twisted structures
on a conical jet at parsec and hundreds of parsec scales
\citep{R89,O89} and apparent motions from subluminal to superluminal
\citep{B99}, and the 3C\,120 jet exhibits subluminal and superluminal
apparent motions and intensity structure that has been interpreted as a
helical twist on a conically expanding jet \citep{G98,W01}.  With the
spatial and time resolution of present observations, we can begin to
discern details such as the trajectory of individual components, the
velocities of components and larger scale structures, and whether there
are similarities between component speeds, trajectories, and larger
scale structures.

Helical structure in relativistic jets can arise as a result of ordered
variation in the flow direction at the central engine, e.g.,
precession, and/or as a result of random perturbations to the jet flow,
e.g, jet cloud interaction \citep{G00}.  Initial perturbations
propagate as waves associated with pinch, helical, elliptical
etc.\ normal mode distortions of the jet.  The helically twisted
structure of a jet is dependent on the initial excited wave frequency
or frequencies and initial amplitudes, and on the subsequent
propagation and growth or damping of these wave frequencies along the
jet.  The resulting helical structure can sensitively depend on jet
speed via the Lorentz factor, on the sound speeds in the jet and
surrounding material, and on the rate of jet expansion.

Only in the past few years have time dependent relativistic
hydrodynamical codes become readily available
\citep{DH94,MMI94,KNM96,FK96,A99a}.  More recently this has allowed
fully 3D relativistic jet simulations to be performed with resolution
sufficient to compare structure with theoretical predictions and/or
with features observed in AGN jets \citep{A99b,A00,H01,HMD02,HH03,A03}.  In
general, computational constraints make it very difficult to simulate
and model the superluminal jets that must lie near to the line of
sight.  Simulations need to be conducted on lengthy computational grids
and store many time steps in order to account for projection and light
travel time effects, e.g., \citet{G97} and \citet{A03}.  This difficulty
can be overcome using theoretical models based on the linearized fluid
equations.

That the the normal wave modes predicted by the theory operate on AGN
jets has been demonstrated by \citet{LZ01} who successfully fitted the
twisted emission threads observed in the 3C\,273 jet by a combination
of helical and elliptical surface and internal normal modes.  Other
similar fitting has been performed by \citet{LHE03}
in the context of M\,87.  Helical twists have been used to explain the
structures in 3C\,120 \citep{G98,G01} and it has been shown that
superluminal motions along a curved trajectory can be explained by
helical jet models, e.g., the radio source 3C~345 \citep{H87,S95}.
While some fitting of helically twisted structure has been performed,
no detailed self-consistent models have been constructed.  Modeling the
emission from relativistic flows associated with the structures
predicted by theory but including all relevant time delays can be used
to provide the proper connection between the observations and the
underlying properties of the outflows.

In this paper I present the basic behavior and appearance of a
helical twist on a conical isothermal constant speed relativistic
jet.  I also show how the appearance of a helically twisted jet can be
used to constrain the macroscopic properties of the jet and external
fluid.

\vspace{-0.7cm}
\section{Applicability of the Linearized HD Equations}

The development of helical jet structure on a relativistic jet can be
obtained using a normal mode analysis of the linearized time dependent
relativistic hydrodynamic equations.  Previous studies indicate that
helical patterns are likely to form in relativistic jets, can induce
significant transverse motion without the development of shocks, and
can produce asymmetries in Doppler boosted emission if jets are
observed at viewing angles on the order of the beaming angle \citep{H00}.

Non-relativistic numerical simulations indicate saturation of the
normal wave modes of jet distortion in certain circumstances.   The
pinch and higher order normal wave modes grow only up to some
saturation amplitude that does not lead to disruption of highly
collimated jet flow \citep{HCR97}. Additionally, short wavelength
helical twisting grows only up to some saturation amplitude that does
not lead to disruption of highly collimated jet flow \citep {XHS00}. In
these cases the normal modes remain within the ``linear'' regime.  On
the other hand, the simulations show that long wavelength helical
twisting serves to disrupt ordered flow when the helical twist grows to
non-linear amplitudes as a result of subsequent filamentation, mass
entrainment and shock formation in the jet.  Disruption by long
wavelength helical twisting can be slowed by high Mach numbers, by
strong magnetic fields, and by jet expansion \citep{RHCJ99,RH00}.

Within the ``linear'' non-destructive regime, helically twisted
structures seen in numerical simulations have been successfully modeled
by theory based on the non-relativistic MHD equations
\citep{HR99,HR02,XHS00} and the relativistic HD equations
\citep{H98,H01,A01,HH03}.  This modeling showed that the theory
correctly models the 3D pressure and velocity fields seen in
non-relativistic and relativistic numerical simulations up to
amplitudes that formally are much larger than linear, i.e., pressure
fluctuations equal to the equilibrium pressure ($P = P_0 \pm P_1$ where
$P_1 \leq P_0$).   Since numerical simulation results indicate jet
disruption by asymmetric structures at non-linear amplitudes, highly
collimated jets must be operating in the linear regime as otherwise
they would be disrupted by asymmetric helical structures.  Thus, I will
assume that pressures and flow fields associated with observed jet
structures can be modeled via the linear theory.

The polarization in jets, except in localized structures, is typically
well below the maximum value of $\sim 70\%$.  This can be taken to mean
that the magnetic field in jets beyond the acceleration region is not
well organized.  Disordered magnetic fields may provide a magnetic
pressure but should not significantly influence the macroscopic normal
mode structures predicted by the relativistic fluid equations.

\vspace{-0.8cm}
\section{Modeling Helical Structures}

Declines in synchrotron intensity along superluminal jets indicate that
$I \propto z_{obs}^{-1.3}$ \citep{Hetal02} where $z_{obs}$ represents the
observed separation distance from the core.  Previous models of similar form
typically have found power-law indices for VLBI jets (e.g.,
\citet{W87}; \citet{UW92}; \citet{X00}) between -1 and -2.  VLA scale
jets show similar power-law indices with increasing jet width,
presumably proportional to the jet radius \citep{BP84}.  If
we assume that this type of power-law decline scales with increasing
jet radius in all cases, then this result strongly argues against adiabatic expansion.  Thus, as a baseline case I will assume constant speed
isothermal expansion.  Isothermal expansion conserves energy flux for a
constant jet speed.  Formally, this means also that the magnetic
pressure is negligible.  It will also be assumed that the material
surrounding the jet is isothermal.  Such an assumption is reasonable if
the jet is surrounded by a cocoon or lobe separating the jet from the
ambient environment.  Here the jet Mach number, $M_{jt} \equiv v_{jt}/a_{jt}$,
remains constant and the external Mach number,  $M_{ex} \equiv v_{jt}/a_{ex}$,
also remains constant with the jet in pressure balance with the
external cocoon or lobe medium.

\vspace{-0.7cm}
\subsection{Helical Twisting: Moving Helical Patterns}

The linearized fluid equations show that a helical twist is
Kelvin-Helmholtz unstable and a helical twist propagates at a wave
speed dependent on jet speed, sound speeds, and on the frequency of the
wave relative to a ``resonant'' or maximally unstable frequency,
$\omega^*$.  For sufficiently supersonic flow, the resonant frequency
is $\omega^* R_{jt}/a_{ex} \sim 1.5$.  The accompanying ``resonant''
wavelength depends on the wave speed.

At very low frequencies relative to the resonant frequency the wave
speed is
\begin{equation}
\eqnum{1a}
v_w = {\gamma^2 \eta \over 1+\gamma^2 \eta }v_{jt}~,
\end{equation}
and the wavelength
\begin{equation}
\eqnum{1b}
\lambda(\omega) = {\gamma^2 \eta \over 1+\gamma^2 \eta }(\omega R_{jt}/v_{jt})^{-1}R_{jt}~.
\end{equation}
The spatial growth length is given by
\begin{equation}
\eqnum{1c}
\ell(\omega) = \gamma \eta^{1/2}(\omega R_{jt}/v_{jt})^{-1}R_{jt}~.
\end{equation}
In the equations above $\gamma = (1 - v_{jt}^2/c^2)^{-1/2}$ is the Lorentz
factor  and $\eta \equiv (a_{ex}/a_{jt})^2$, where $a_{ex}$ and $a_{jt}$
are the sound speeds outside and inside the jet.  The sound speed
$a\equiv [ \Gamma P_0 / (\rho _0+[\Gamma /(\Gamma-1 )]P_0/c^2) ]
^{1/2}$, where $4/3\leq \Gamma \leq 5/3$ is the adiabatic index.  The
density, $\rho_0$, and pressure, $P_0$, are measured in the proper
fluid frames, and since pressure balance has been assumed $\eta$ is
effectively an enthalpy ratio.  For constant $\omega << \omega^*$ the
wave speed and growth length remain constant as the jet expands.

The wave speed changes as resonance is approached and at resonance the wave speed is (Hardee 2000)
\begin{equation}
\eqnum{2a}
v_w^* \approx \frac{\gamma
[M_{jt}^2-v_{jt}^2/c^2]^{1/2}}{[M_{ex}{}^2-v_{jt}^2/c^2]^{1/2}+\gamma
[M_{jt}^2-v_{jt}^2/c^2]^{1/2}}~v_{jt}~. 
\end{equation}
The corresponding ``resonant'' wavelength is
\begin{equation}
\eqnum{2b}
\lambda^* \approx 3.75 \frac{\gamma
[M_{jt}^2-v_{jt}^2/c^2]^{1/2}M_{ex}}{[M_{ex}{}^2-v_{jt}^2/c^2]^{1/2}+\gamma
[M_{jt}^2-v_{jt}^2/c^2]^{1/2}}R_{jt}~,
\end{equation}
and the ``resonant'' spatial growth length is
\begin{equation}
\eqnum{2c}
\ell^* \gtrsim \gamma M_{jt} R_{jt}~.
\end{equation}
This minimum growth length is typically greater than a straight
extrapolation of the low frequency result to the resonant frequency.
At very high frequencies and short wavelengths relative to resonance
the wave speed becomes
\begin{equation}
\eqnum{3}
v_w \approx \frac{v_{jt} - a_{jt}}{1 -
a_{jt}v_{jt}/c^2} 
\end{equation}
and the wave propagates like a backwards moving sound wave in the
reference frame of the jet fluid.  The growth rate is vanishingly small
in the high frequency and short wavelength limit.

The analytic expressions can be used to provide limiting estimates of
wave speed, wavelength and amplitude growth to be expected of a fixed
frequency wave along an expanding jet.  However, detailed development
requires numerical solutions to the dispersion relation describing
helical wave propagation and growth.  Solutions to this dispersion
relation for a jet speed of $\beta_{jt} \equiv v_{jt}/c = 0.99$, an adiabatic
index of $\Gamma = 13/9$ appropriate to a mixture of hot leptonic and
cold baryonic material \citep{S57}, and representative choice of sound
speeds are shown in Figure 1.  Panels in Figure 1 have been arranged so
that the resonant frequency normalized by the jet speed and jet radius,
$\omega^*R_{jt}/v_{jt}$, remains nearly constant along a diagonal direction.
Similar wave speed and wavelength behavior as a function of $\omega
R_{jt}/v_{jt}$ occurs along the vertical direction. For a sufficiently
relativistic jet, behavior is governed entirely by the Lorentz factor
and sound speeds.

\begin{figure}[p]
\vspace{17.5cm}
\includegraphics{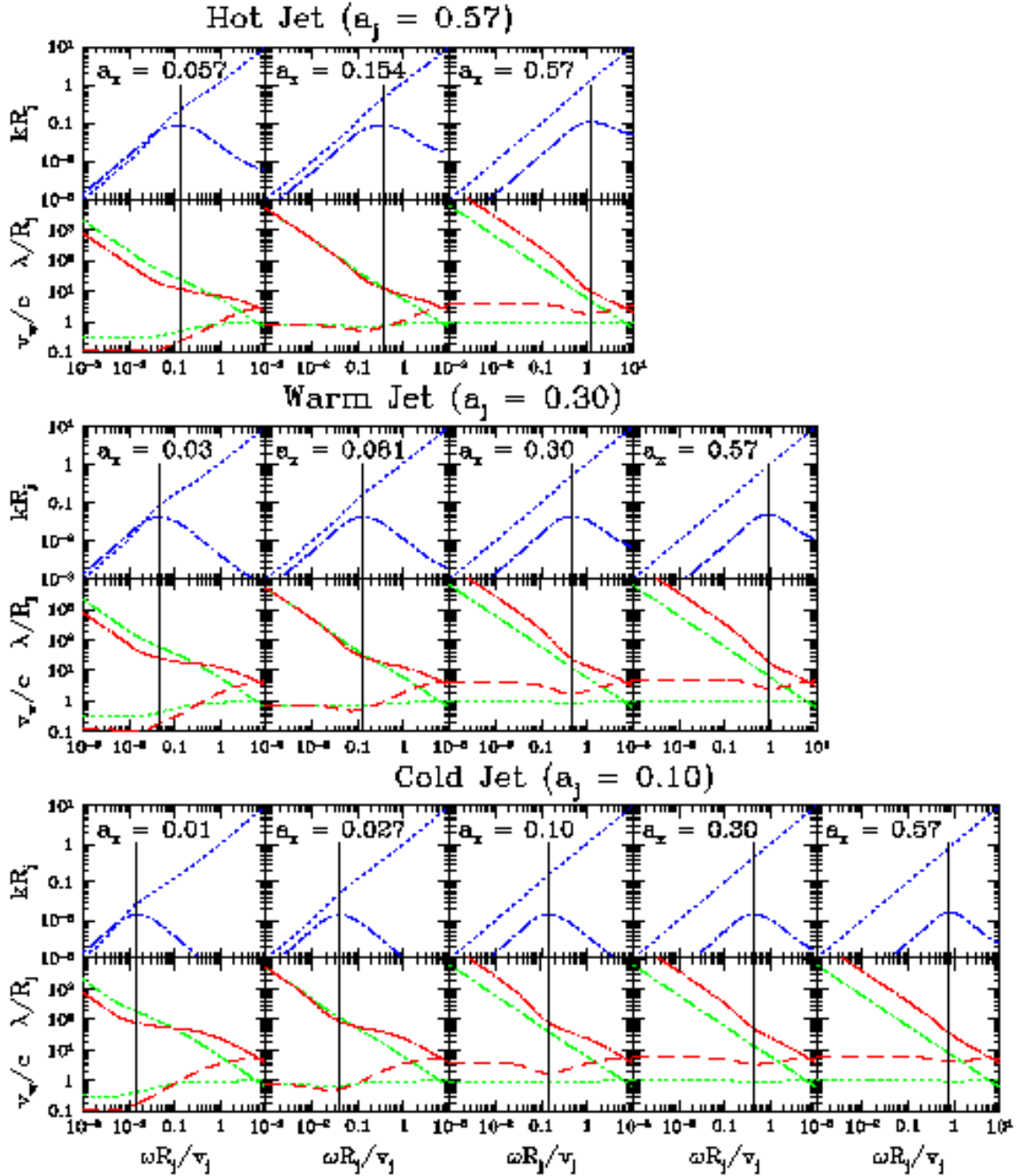}
\caption {\footnotesize \baselineskip 10pt  Solutions to the dispersion
relation $kR_{jt}$ as a function of $\omega R_{jt}/v_{jt}$ for a helical wave.
\Blue{Dotted} (\Blue{Dash-dot}) lines indicate the real (imaginary) part of the wavenumber, and the sound speed is indicated in units of c.  The
intrinsic (\Green{dotted line}) and apparent (\Red{dashed line}) wave
speeds and intrinsic (\Green{short dash-dot line}) and apparent
(\Red{long dash-dot line}) wavelengths are shown immediately below the
appropriate dispersion relation solutions. The vertical line indicates
the location of the resonant frequency.
\label {fig1}}
\end{figure}

Figure 1 also shows the intrinsic and apparent wave speeds and
wavelengths associated with the dispersion relation solutions where I
have chosen a jet viewing angle of $\theta_0 = 14\arcdeg$.   With this
assumed viewing angle and a flow Lorentz factor of $\gamma \equiv [1
- \beta_{jt}^2]^{-1/2} \sim 7$, the maximum apparent superluminal speed is
$\beta^{obs}_{max}\approx 6$.  Note that at this viewing angle
subluminal apparent motions occur for wave speeds $\beta_w \equiv v_w/c
< 0.78$ with $\gamma_w < 1.6$.  The apparent helical wavelength in the
observer frame is related to the intrinsic wavelength by
$\lambda^{obs}=\beta^{obs}_w \lambda$ and 
the intrinsic wavelength can be greater or less than the observed
wavelength depending on wave speed and viewing angle.  For our viewing
angle, a wave speed equal to the flow speed implies $\lambda^{obs}
\sim 6 \lambda$ but when $\beta_w < 0.78$ $\lambda^{obs} < \lambda$.
When sound and jet speeds remain constant, a helical wave of fixed
frequency, $\omega$, remains at constant wavelength in the long and
short wavelength regimes where the wave speed is constant.  

Even with constant sound and jet speeds, a helical wave of fixed
frequency changes propagation speed and wavelength as $\omega R_{jt}/a_{ex}$
moves from the low frequency to the high frequency regime as a result
of increase in $R_{jt}$. 
Change in the observed wavelength is amplified relative to
change in the intrinsic wavelength, $\lambda(z) = [\beta_w(z)/\beta_w(z_0)] \lambda(z_0)$, with
\begin{equation}
\eqnum{4}
\lambda^{obs}(z\sin\theta_0) = { \sin\theta_0 \over
1- \beta_w(z) \cos\theta_0}{\beta_w(z) \over \beta_w(z_0)} \lambda(z_0)~.
\end{equation}
In equation (4), $z$ is the intrinsic distance along the jet axis where
$z_{obs} \equiv z\sin\theta_0$ is the observed distance.  The
analytical and numerical solutions to the dispersion relation indicate
that significant change in the wavelength can occur when $a_{ex}/a_{jt}
< 1$.  Much reduced variation occurs in observed wave speed when
$a_{ex}/a_{jt} > 1$. In particular, note that $a_{ex}/a_{jt} \ll 1$
implies $\beta_w \ll 1$ when $\omega R_{jt}/v_{jt} \ll \omega^*
R_{jt}/v_{jt}$ and $\beta_w \lesssim 1$ when $\omega R_{jt}/v_{jt} \gg
\omega^* R_{jt}/v_{jt}$ (leftmost panels in Figure 1).  The resulting
broad plateau seen in $\lambda^{obs}/R_{jt}$ implies that a fixed
frequency helical wave propagating along an expanding jet will appear
to increase in wavelength approximately proportional to $R_{jt}$ over
the corresponding normalized frequency range, and will transition from
less than to greater than the intrinsic wavelength.  The corresponding
normalized frequency range allows this behavior to be observed over up
to two orders of magnitude of jet expansion.  This wavelength change
would be accompanied by a similar large change in the observed wave
speed.

As a helical twist is Kelvin-Helmholtz unstable, the displacement
amplitude of the jet surface grows according to
$$
A = A_0~exp \left[ \int_{z_0}^{z}\ell(z)^{-1} dz \right]~,
$$
where $A_0$ is the displacement amplitude at $z_0$.
However, the magnitude of accompanying velocity and pressure
fluctuations grows according to $A(z)/R_{jt}(z)$, and for constant jet expansion
$$
{A(z) \over R_{jt}(z)} = {A_0 \over R_0}~{exp \left[ \int_{z_0}^{z}\ell(z)^{-1} dz \right] \over
\left[ 1 + (z - z_0)\psi/R_0 \right]}~,
$$
where $R_0$ is the jet radius at $z_0$ and  $R_{jt} = R_0 + (z -
z_0)\psi$.  In the absence of non-linear stabilizing effects, $\ell(z)$
from equation (1c) in the low frequency limit is constant, begins to
increase as resonance is approached and then rapidly increases at
higher frequencies.  For a jet to both exhibit helical twisting and
remain collimated the amplitude must grow sufficiently rapidly but not
so rapidly that large amplitudes are reached while the wave is in the
low frequency regime.  In order that the helical twist not be damped in
the low frequency regime we must have that $A(z)/R_{jt}(z) \ge
A_0/R_0$.  Scaling of the growth length proportional to the wavelength
in the low frequency regime coupled to a minimum in the growth length
at $\omega^* R_{jt}/a_{ex}$ means that the maximum in $A(z)/R_{jt}(z)$
occurs when $\omega R_{jt}/a_{ex} > \omega^* R_{jt}/a_{ex}$.  Note
however that constant pressure and velocity fluctuations can be
maintained for decreasing $A(z)/R_{jt}(z)$ when $\omega R_{jt}/a_{ex}
>> \omega^* R_{jt}/a_{ex}$ (see \S 3.2).

Theoretically, these basic properties allow estimation of jet
parameters, i.e., sound speeds for a given Lorentz factor and/or
viewing angle without detailed modeling of the intensity structure.
The viewing angle and flow Lorentz factor are constrained by the
highest observed superluminal speed, assumed related to the flow speed.
Lack of damping sets an upper limit to $\gamma M_{jt}$.  A constant
wavelength indicates a helical pattern in the low or high frequency
limits. In the low frequency limit this sets a value for
$\gamma a_{ex}/a_{jt}$.  In the high frequency limit this sets a value for
$a_{jt}$.  If the helical wavelength appears to scale approximately with
the jet radius over some expansion range, then this can be used to set
an upper limit to $\gamma a_{ex}/a_{jt}$ and lower limit to $a_{jt}$ from lower
and upper limits to the wave speed, respectively.

\vspace{-0.7cm}
\subsection{Helical Twisting: Decoupling the Pattern and Fluid Flow}

Examination of the fluid flow accompanying a growing helical twist on
an expanding jet can be accomplished by constructing 3D theoretical
data cubes using the expressions giving the pressure and velocity
structure associated with the normal modes \citep{H98,H00}.  Formally, these expressions assume a cylindrical jet.  However,
conical jet expansion at small opening angle does not significantly
modify local solutions to the dispersion relation \citep{H84}.  In
what follows I adopt a jet half opening angle of $\psi = 0.025$ radian
that is less than the relativistic Mach angle, $\sim (\gamma
M_{jt})^{-1}$, for the two cases to be considered.  The flow and pressure
field accompanying a helical twist can be evaluated quantitatively by
taking 1-D cuts through the theoretically generated 3-D data cubes at
constant values of $x/R_{jt}$ as illustrated in Figure 2.  For 1-D cuts in
the x-z plane $v_x$ and $v_y$ are radial and azimuthal velocity
components in cylindrical coordinates.
\vspace {4.5cm}
\begin{figure}[h!]
\includegraphics{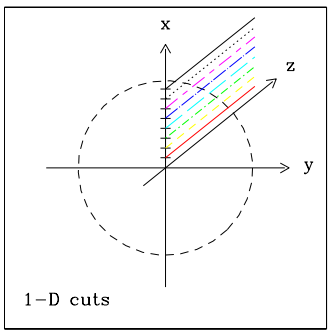}
\caption{\footnotesize \baselineskip 8pt 
1-D cuts at x/R = 0.11 (\Red{solid line}), 0.22 (\Yellow{short dash line})
0.33 (\Green{dot-short dash line}), 0.44 (\Cyan{long dash line}), 0.55
(\Blue{dot-long dash line}), 0.66 (\Magenta{short \& long dash line}), 0.77 (dotted line), \& 0.88 (solid line). Here we look downstream and $z$ is into
the page.
\label{fig2}}
\end{figure}

The evolution of the pressure and velocity components of frequencies
$\omega R_0/v_{jt} =$ 0.01, 0.03, 0.1 for a ``hot'' jet ($a_{jt} = 0.57$)
with $a_{ex} = 0.057$ and for a ``warm'' jet ($a_{jt} = 0.30$) with $a_{ex} =
0.081$ is shown in Figure 3. These two cases, see Figure 1, have been
chosen because the resonant frequency and intrinsic wavelengths are
comparable but sound speeds are significantly different.  As a result,
the wave speed on the warm jet is significantly larger and growth rate
on the warm jet is significantly reduced compared to the hot jet. In
order to obtain similar growth with ``saturation'' amplitude occurring
at a frequency $\omega >> \omega^*$, an initial amplitude of $A_0/R_0
=$ 0.0007 and 0.07 is chosen for hot and warm jets, respectively.  The
large difference in initial amplitudes indicates an extreme sensitivity
of the growth rate to the jet sound speed, i.e., decrease in jet sound
speed by a factor 2 has required an initial amplitude increase by a
factor 100. In fact $A/R_{jt}$ initially declines for the low frequency
wave on the warm jet.   Saturation is imposed when pressure
fluctuations reach the maximum allowed by the linear analysis.
Saturation was imposed at the medium frequency on hot and warm jets and
at the high frequency on the warm jet. It was found that saturation in
pressure fluctuations was maintained by imposing a decrease in the
displacement amplitude of the form $A = A(z) \times (2k_R^*/k_R)^n$
when $(2k_R^*/k_R) \le 1$, where $k_R$ is the real part of the
wavenumber, $k_R^*$ corresponds to resonance and $n \le 2$.
\begin{figure}[p]
\vspace{19.0cm}
\includegraphics{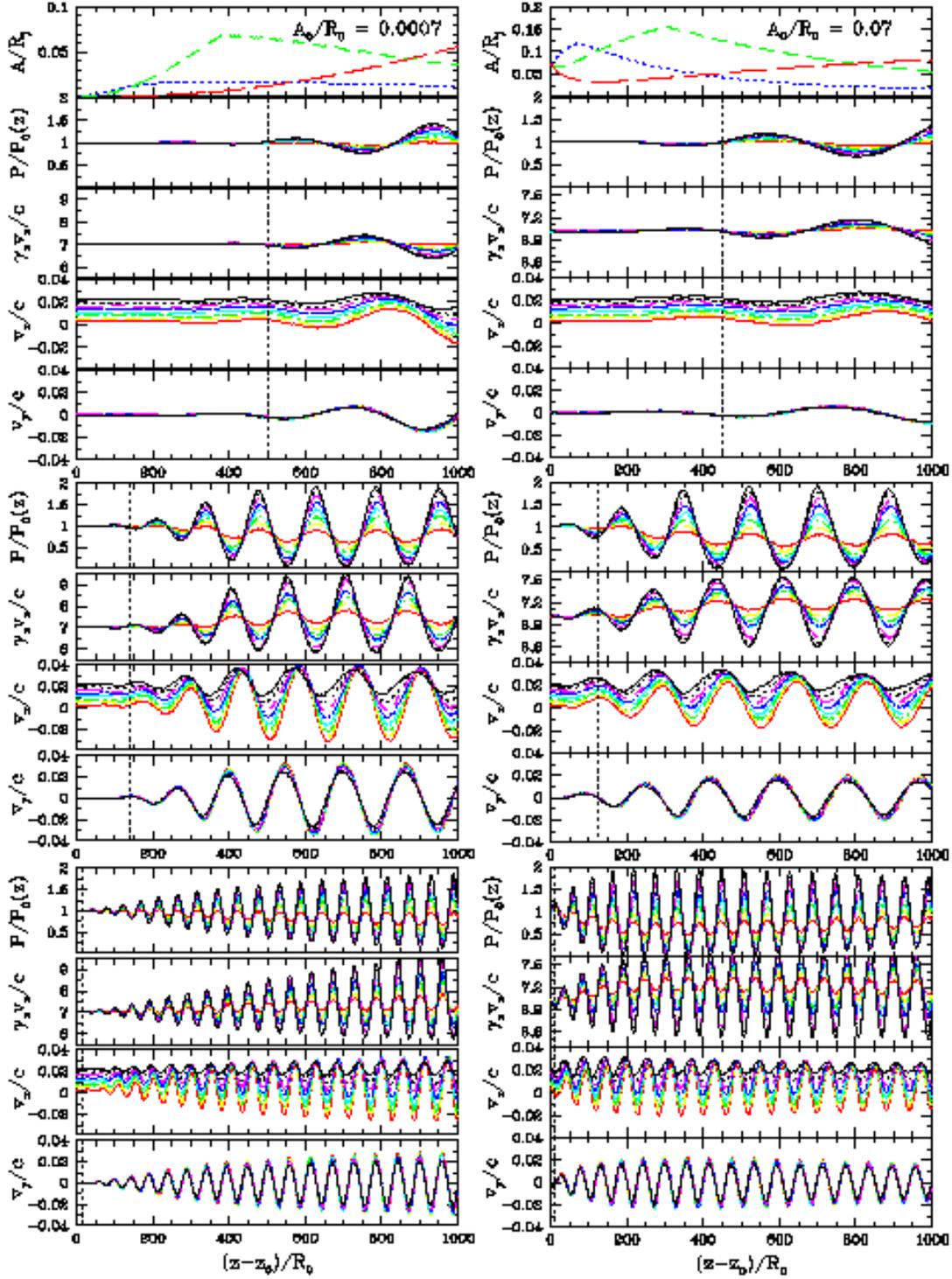}
\caption {\footnotesize \baselineskip 10pt The leftmost (rightmost) set
of panels is for the hot (warm) jet.  The topmost panel shows the
amplitude growth of $A/R_{jt}$ as a function of distance for low
(\Red{long dash line}), medium (\Green{short dash line}) and high
(\Blue{dotted line}) frequencies. Three
sets of panels below show the normalized pressure $P/P_0(z)$, and
velocity components $\gamma_z v_z/c$, $v_x/c$ and $v_y/c$ for low (top
set), medium (middle set) and high (bottom set) frequency.  The dotted
vertical line in the panels indicates the position at which the wave
frequency equals the local resonant frequency.
\label {fig3}}
\end{figure}

The velocity fluctuations on the warm jet are $\approx$ 55\% of that on
the hot jet for a given pressure fluctuation and the difference
reflects the decrease in sound speed and increase in wave speed on the
warm jet.  Note that jet expansion is indicated by the spread in radial
velocities across the jet.  While not readily apparent from these
plots, the jet attempts to move bodily around the central axis.  See,
for example, the velocity vectors in Figures 4 \& 7 in \citet{H01} that
indicate uniform sideways motion of the jet fluid across a jet cross
section.  Here sideways motion is impeded at the jet surface by the
external fluid.  The ``heavier'' external fluid surrounding the hot jet
impedes the surface motion, relative to motion of the jet interior,
more than is the case for the warm jet. Note that shorter intrinsic
wavelengths are associated with higher wave speed in these two cases.
For a higher wave speed the fluid does not need to move as rapidly in
the transverse direction to circulate through the helical pattern.
This effect tends to  cancel the higher circulation speed otherwise
required by a shorter wavelength helical pattern.

The appearance of the high pressure ridge corresponding to the low,
medium and high frequency fluctuations shown in Figure 3 is shown in
Figure 4.  The images in Figure 4 are line of sight integrations
through 3-D pressure data cubes.  In constructing the data cubes it is
assumed that the pressure on the unperturbed conical jet declines as
$P_0(z) \propto R_{jt}(z)^{-2}$.
\begin{figure}[h]
\vspace{6.5cm}
\includegraphics{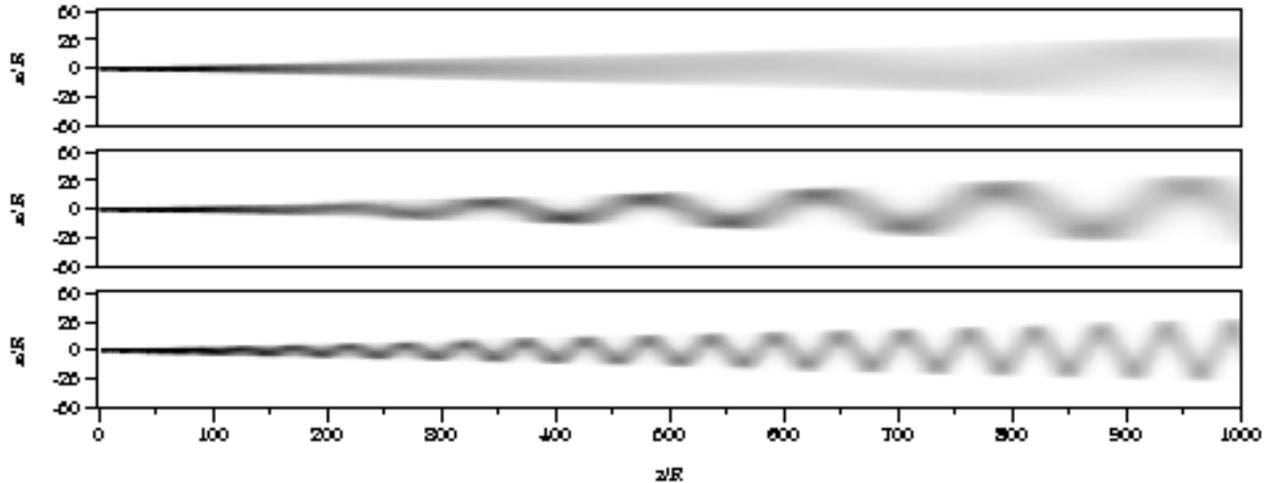}
\caption {\footnotesize \baselineskip 10pt Line of sight integration through 3-D
pressure data cubes.  The three panels show the appearance of the
helical pressure ridge that accompanies the low (top), medium (middle)
and high (bottom) frequency hot and warm jet fluctuations shown in
Figure 3.
\label {fig4}}
\end{figure}
Only one set of images for low, medium and high frequencies is shown as
hot and warm jets appeared nearly identical.  For all frequencies the
surface displacement of the jet is small and the images show that the
high pressure ridge appears to lie within the conical jet's surface.
In general, the jet fluid does not follow the path of the high pressure
ridge.  If we define the pitch angle of the flow by $tan~\theta_f =
(v_\perp /v_z)$, then $tan~\theta_f \le$ 0.035 \& 0.02 for saturation
amplitudes on the hot and warm jets, respectively, where $v_\perp /v_z
\sim \mid v_x/c \mid_{max} \sim  \mid v_y/c\mid_{max}$ from Figure 3.
This maximum flow pitch angle is on the order of half the relativistic
Mach angle, i.e., $\theta_f \le \theta_{rel}/2 = (2\gamma
M_{jt})^{-1}$.  The helical pitch of the high pressure ridge  defined
by $tan~\theta_h = 2\pi R_{jt}/\lambda$ varies from $0.04 <
tan~\theta_h < 1$ at the medium frequency and $0.11 < tan~\theta_h < 3$
at the high frequency as $0 \le (z - z_0)/R_0) \le 1000$ on both hot
and warm jets.   As a result of the decoupling between the pattern and
fluid flow, the variation in flow angle, $\theta = \theta_0 \pm \Delta
\theta$ where $\Delta \theta \lesssim \theta_f$, is always less than
would be inferred from the helical pitch of the high pressure ridge.

\vspace{-0.7cm}
\subsection{Helical Twisting: Observing Helical Patterns}

The brightest radio emission is expected to trace the path of the high
pressure ridge.  In constructing line of sight images it will be
assumed that a pseudo-synchrotron emissivity at fixed frequency can be
written as
\vspace{-0.2cm}
$$
\epsilon_\nu \propto n_{jt}^{1 - 2 \alpha}p_{jt}^{2\alpha}(B~sin~\theta_B)^{1 +
\alpha}D^{2 + \alpha}
\vspace{-0.2cm}
$$ 
where $\theta_B$ is the angle of the magnetic field to the line of
sight and $D \equiv [\gamma(1 - \beta~cos~\theta)]^{-1}$ is the Doppler
boost factor for fluid flowing with speed $\beta$ at angle $\theta$ to
the line of sight.   I have used $\alpha = 0.5$ where $\nu^{-\alpha}$
and set $\theta_B =$~constant.  This form for the pseudo-synchrotron
emissivity is necessary when the synchrotron emitting particles are not
explicitly tracked \citep{CNB89}.  That pseudo-synchrotron intensities
provide reasonable images of internal jet structures is indicated by
the work of \citet{JRE99}.  Along the conical constant velocity
isothermal jet the particle number density and pressure are assumed to
decline $\propto R_{jt}^{-2}$.  The magnetic field strength is assumed to
decline as $B \propto n_{jt}^{2/3} \propto R_{jt}^{-4/3}$ appropriate to a
disordered magnetic field \citep{HB00}.

\vspace{-0.7cm}
\subsubsection{Individual Frequencies}

Line of sight pseudo-synchrotron images corresponding to the low,
medium and high frequency hot and warm jet cases shown in Figures 3 \&
4 are shown in Figure 5.  In these images all wave motion, light travel
time and Doppler boosting effects appropriate to a jet at a $\theta_0 =
14^\circ$ viewing angle have been incorporated, and the apparent
opening angle of the conical jet is $\approx 12\arcdeg$. The
$x$,$y$,$z$ coordinate system is defined by the jet with $x$ in the sky
plane and $z$ down the jet axis with the hypothetical observer located
in the $y-z$ plane.
\begin{figure}[h]
\vspace{9.7cm}
\includegraphics{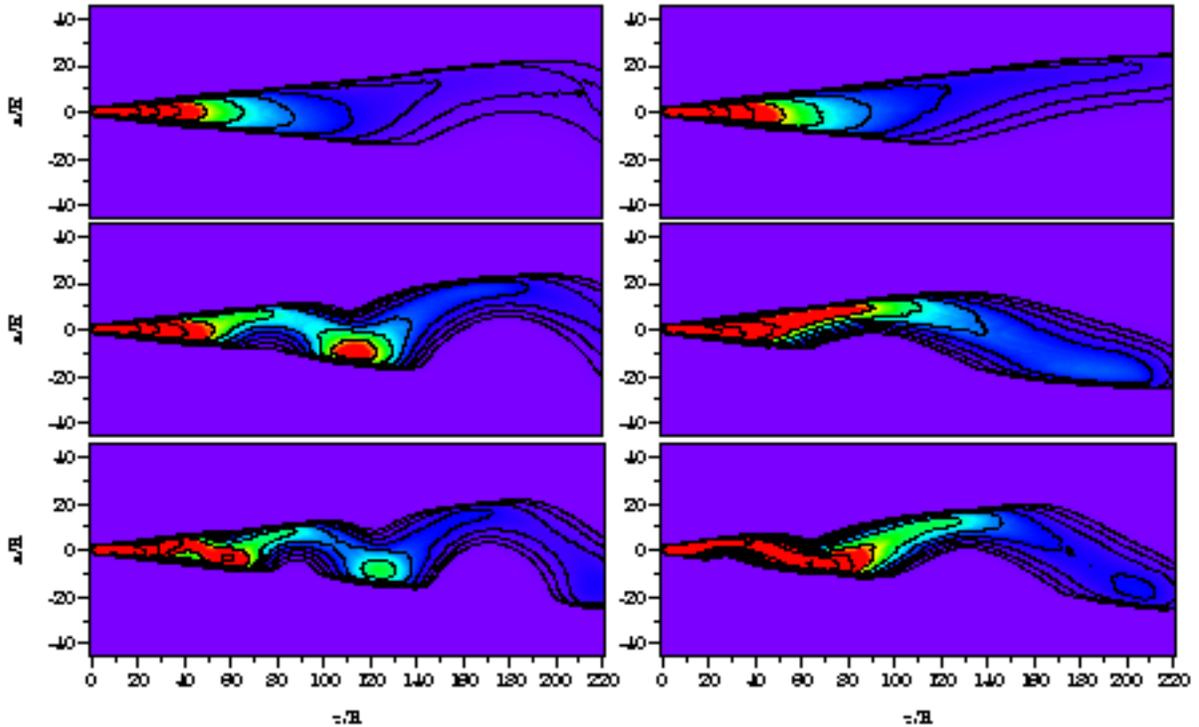}
\caption {\footnotesize \baselineskip 10pt  Pseudo-synchrotron
intensity images for low (top), medium (middle) and high (bottom)
frequency helical waves on hot (left panels) and warm (right panels)
jets at a viewing angle of $\theta = 14 \arcdeg$.  The contours levels
are in factors of 2.
\label {fig5}}
\end{figure}
Comparison between Figure 4 (similar to a
pseudo-synchrotron intensity image of the jet in the plane of the sky)
and Figure 5 provides a striking example of the effect of wave motion
and the accompanying light travel time effects on the apparent
wavelength.  On the hot jet the apparent wavelength varies from less
than to greater than the intrinsic wavelength from low to high
frequencies, respectively.  On the warm jet the apparent wavelength
varies from comparable to much greater than the intrinsic wavelength
from low to high frequencies, respectively.  At all three frequencies
the longer apparent wavelength on the warm jet is the result of faster
wave motion.  The reduced wave speed on the hot jet and accompanying
reduced apparent wavelength are sufficient to have produced intensity
knots.  Increase in the apparent wavelength with $z$ at medium and high
frequencies on the hot jet and at high frequency on the warm jet
results from increase in the intrinsic wave speed.

The existence of intensity knots and ratio of brightnesses between knot
and interknot regions can be used to constrain helical flow models,
e.g., \citet{W01}.  For the hot jet medium and high frequency
helical twists, the intensity ratio between knot and interknot regions
is $\approx 4$. In general, such intensity knots are the result of a
combination of helical twist projection and Doppler boosting effects
where the magnitude and direction of flow is modified by high and low
pressure regions.  For nearly constant Lorentz factor, the Doppler boost
factor is approximately given by
$$
D = D_0 \pm \Delta D \approx D_0\left[ 1 \pm {\beta_0 \theta_0 \over (1 - \beta_0 cos~\theta_0)}~\Delta \theta \right]
$$
where $\Delta \theta \lesssim \theta_f$.  Along the hot conical jet
with a saturated helical twist at $\omega > \omega^*$ the near side can
be boosted relative to the far side and the bottom can be boosted
relative to the top with $\theta_f \le 0.035$~radian. Thus, with $D_0
\sim 3.6$ we expect $\Delta D \sim 0.2D_0 \lesssim 0.7$ and a maximum
variation in $D^{2.5}$ of $\sim 2.5$.  About half of the
intensity variation results from simple projection effects and half
from Doppler boosting effects.  The higher wave speed associated with
the warm jet has reduced transverse motion and Doppler boosting
contrast, and the longer apparent wavelength has reduced projection
effects such that intensity knots are almost eliminated.

\vspace{-0.7cm}
\subsubsection{Multiple Frequencies}

For a single frequency excited by precession of the central engine the
behavior of a helical twist is predictable over many orders of
magnitude of jet expansion.  Difficulties arise if multiple frequencies
are excited.  At any location along the ``linear'' expanding jet the
frequency that has grown the most will be above the local resonant
frequency.  This frequency reaches saturation or some lesser amplitude
depending on the initial amplitude and growth rate, but could be
overwhelmed by a helical wave at lower frequency that continues to grow
and so on.  Since the resonant wavelength scales with the jet radius,
we would expect to see approximate scaling with the jet radius, albeit
with jumps in wavelength.

Perhaps the largest uncertainty in dealing with multiple frequencies
lies in the interaction between multiple frequencies associated with
the same wave mode.  Numerical simulations have shown that the faster
growing higher order normal modes (elliptical, triangular etc.) do not
slow the growth of a helical mode wave and decline in amplitude as the
longer wavelength helical twist grows \citep{HCR97}.  Another
simulation has shown that pinch mode waves show relatively abrupt
change to longer wavelength on a conically expanding axisymmetric jet
\citep{A01}.  Therefore I follow the simple assumption that a high
frequency saturated helical wave will decline in amplitude as a lower
frequency helical wave grows as specified by the linear theory.
Specifically I consider the hot jet with a combination of low and
medium frequency waves and with a combination of low and high frequency
waves.  I choose the hot jet as the variation in wave speed from low to
high frequencies is more extreme than on the warm jet.  Jet speed and
opening angle remain the same as the previous single frequency models.

The results of this study are shown in Figure 6.  Figure 6 contains
plots of $A/R_{jt}$ for each wave, 1-D cuts of the combined wave
pressure and velocity fluctuations, and pseudo-synchrotron intensity
images for the jet in the plane of the sky and for the jet at a viewing
angle of $14 \arcdeg$.  The combined pressure and velocity fluctuations
are the result of a linear combination of fluid displacements
associated with the individual waves from which the velocity components
and pressure are computed.  Note that  the pressure is not a linear
combination of pressures associated with the individual waves. Combined
low and medium frequency waves grow from an initial $A_0/R_0 =
0.0007$.  Saturation is reached for low and medium frequency waves when
the individual wave frequency is somewhat larger than twice the local
resonant frequency.  The linear combination of the two waves remains
saturated if the amplitude of the individual waves behaves like $A
\times (2k^*/k)^3$ when $(2k^*/k) \le 1$.  Previously, I found
$(2k^*/k)^2$ maintained saturation for the medium frequency wave
alone.  For combined low and high frequency waves the high frequency
wave grows from an initial $A_0/R_0 = 0.001$ in order to reach
saturation before declining in amplitude like $A \times (8k^*/k)^3$
when $(8k^*/k) \le 1$.
\begin{figure}[h!]
\vspace{14.3cm}
\includegraphics{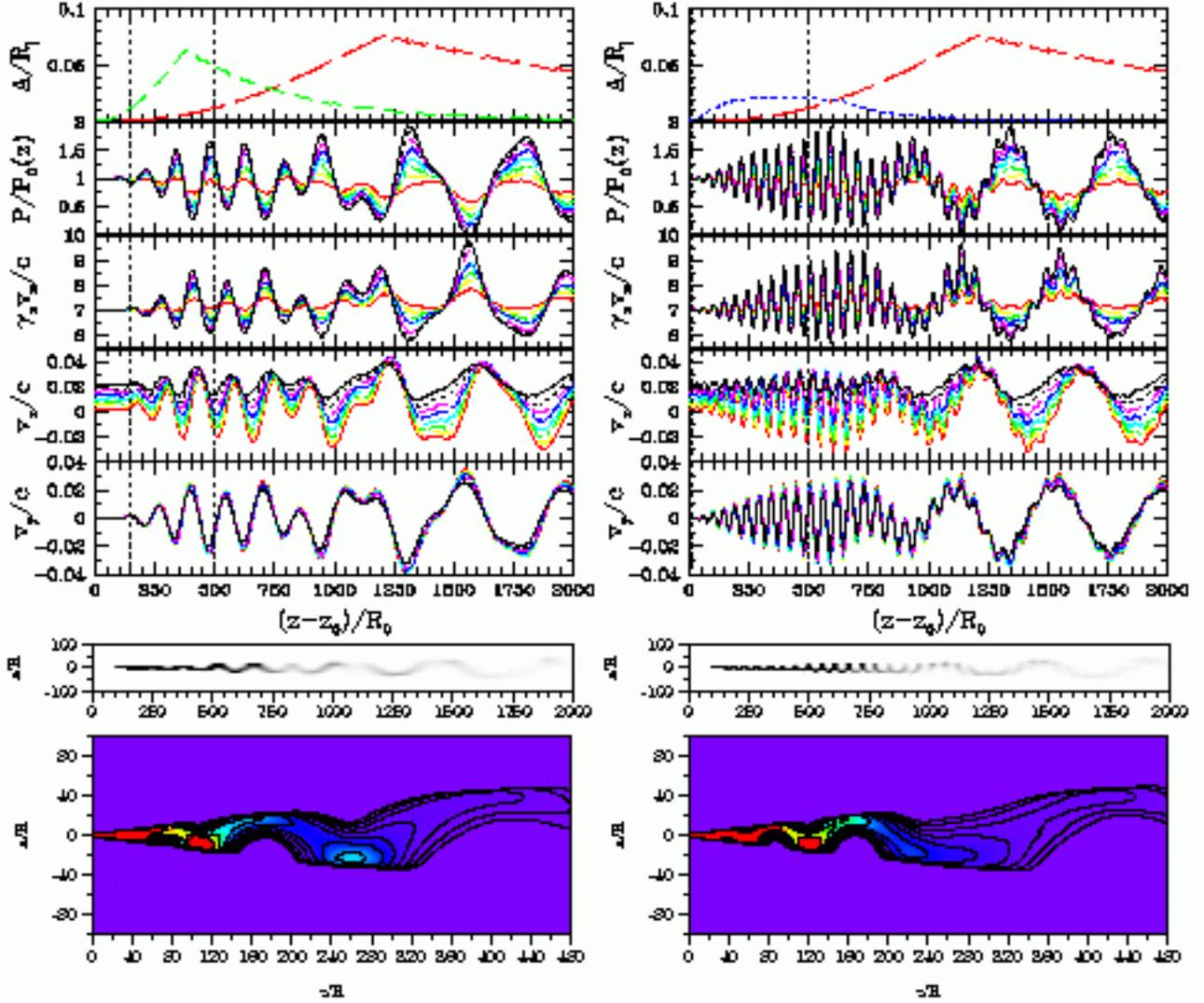} 
\caption {\footnotesize \baselineskip 10pt The leftmost (rightmost) set
of panels is for the low/medium (low/high) frequency combinations.  The
topmost panel shows the amplitude growth of $A/R_{jt}$ as a function of
distance for low (\Red{long dash line}), medium (\Green{short dash
line}) and high (\Blue{dotted line}) frequencies.  Panels below show
the pressure and velocity components as in Figure 3.  The two dotted
vertical lines in the panels indicate the position at which the higher
and lower wave frequency equals the local resonant frequency.  Below
these panels pseudo-synchrotron intensity images for low/medium and
low/high frequency combinations for the jet in the plane of the sky and
for the jet at a $14 \arcdeg$ viewing angle are shown.  The contour
levels are in factors of 2.
\label {fig6}}
\end{figure}
  
The rapid motion of the high frequency wave leads to significant light
travel time effects across the jet diameter, visible as cusps at the
top of the jet, for the image corresponding to the jet in the plane of
the sky.  At the $14 \arcdeg$ viewing angle relatively smooth scaling
of the helical wavelength with jet radius (low plus medium frequency
wave) or more abrupt jump in helical wavelength (low plus high
frequency wave) is observed.  Knot motion can be complex for multiple
frequency waves.  For the low and medium frequency combination the knot
at $z/R \sim 110$, $(z - z_0)/R_0 \sim 450$, will move with
$\beta_w^{obs} \sim 0.35$.  The knot at $z/R \sim 260$, $(z - z_0)/R_0
\sim 1070$, is a combined feature and could show a component from the
medium frequency wave with $\beta_w^{obs} \sim 1$ moving through a
slower $\beta_w^{obs} \sim 0.3$ feature associated with the low
frequency wave.  For the low and high frequency case the knot at $z/R
\sim 120$ [$(z - z_0)/R_0 \sim 500$] will move with $\beta_w^{obs} \sim
2$  while the low frequency wave moves with $\beta_w^{obs} < 0.3$.  At
an apparent distance of $z/R = 500$ [$(z - z_0)/R_0 \sim 2100$] the low
frequency wave moves with $\beta_w^{obs} \sim 0.6$.  Recall that knots
moving with the flow would appear to move at $\beta_f^{obs} \sim 6$ and
the maximum possible wave speed found using equation (3) is
$\beta_w^{obs}\mid_{max} = 3.7$.

\vspace{-0.7cm}
\section{The inner 3C\,120 jet as an example}

The radio source 3C~120 exhibits structure on scales from under a
parsec to hundreds of kiloparsecs \citep{W87}. The galaxy has a
redshift $z=0.033$ \citep{B80}, and following \citet{G01} I let $H_o =
h~ 100$~km~s$^{-1}$~Mpc$^{-1}$ with $h = 0.65$ and there are $\approx
0.70 h_{65}^{-1}$~pc~mas$^{-1}$. Recent high frequency VLBI
observations at 86, 43 and 22 GHz \citep{G00,G01} have produced high
resolution images of the inner jet.   Components $s, r, m, o, h~\&~d$,
shown here in Figure 7 (see also \citet{G01} Fig.\ 1), located
successively farther from the core have motions $\approx$~0.27, 0.40,
0.49, 1.83, 1.75 \& 1.71~\masr\, respectively, where 1~\masr $= 2.34
h_{65}^{-1}$ c. The separation of these components increases from $s$
to $o$ with $o$ at $\approx 2$~mas from the core.

\begin{figure}[h]
\vspace{9.5cm}
\includegraphics{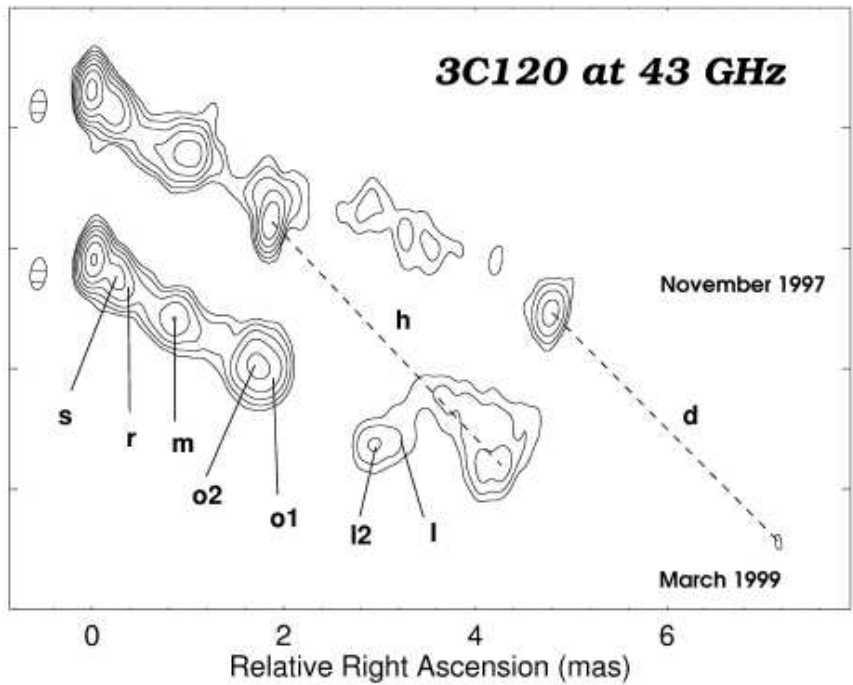}
\caption {\baselineskip 13pt Synchrotron intensity images at 43 GHz of
the 3C\,120 jet at two epochs. Component o in the text corresponds to
o2 \& o1 in the image. Contour levels are in factors of 2.  Image
provided by Jose-Luis G\'omez. 
\label {fig7}}
\end{figure}

\citet{G98} suggested that components $h$ and $d$ located at $\sim
2$~mas and $\sim 5$~mas from the core in 43 and 22 GHz images in
\citet{G99} (Figure 1) were the result of a projected helical path and
experienced a different evolution from other components that formed a
curved path between and beyond these two components.  The relatively
constant spacing and motion along with ``saturated'' component
intensity structure beyond 2~mas indicates that these components
represent a helical twist in the high frequency regime. In \citet{G01}
fast moving components between $o$ and $h$ (e.g., l \& l2 in Fig.\ 7) show
motions from 2.17 to 2.48~\masr.  Similar fast motions, $\sim$
2.2~\masr\ to 3.2~\masr\, are found by \citet{W01} at 5~GHz out to
14~mas from the core.  An average apparent motion associated with fast
moving components is $\sim 2.6$~\masr.  The resulting apparent
superluminal motion, $\beta^{obs} \sim 6$, constrains the viewing angle
to less than $19 \arcdeg$ and a minimum fast moving component speed of
$\beta = 0.986$ is obtained at a viewing angle $\theta = 9.6 \arcdeg$.
It is likely that the average speed of the faster moving components is
indicative of the flow speed, and the motion of components $h$ and $d$
is indicative of helical wave motion in the high frequency regime.

Components $s - m$ have been interpreted as pinch structures
\citep{G01} triggered by passage down the jet of a shock associated
with component $o$.  Such an effect was observed by \citet{A01} in
numerical simulation.  In what follows I will assume instead that these
newer components are also associated with a helical wave and that the
average apparent speed, $\beta^{obs} =$~4.14 c, of components $o,
h~\&~d$ represents motion of the helical wave in the high frequency
regime.  The apparent wavelength in the high frequency regime is then
given by the spacing, $2.5 - 3$~mas, between components $o, h~\&~d$.
The apparent increasing speed of components $s, r, m~\&~o$ is
associated with a change in helical wave speed as a wave of single
frequency makes a transition from the low to high frequency regime by
virtue of jet expansion.  The apparent speed of component $s$ nearest
to the core, $\beta^{obs} =$~0.63, is assumed to represent motion of
the helical wave in the low frequency regime.

In what follows I will consider a single viewing angle of 9.6\arcdeg.
Obviously a range of angles and computation of synchrotron intensity is
needed to constrain the viewing angle but this modeling is beyond the
scope of the present work.  With an observed jet opening angle of $\sim
12\arcdeg$, the true half-opening angle becomes $\approx 1\arcdeg$.
At the assumed viewing angle the true wave speed changes from low to high
frequency regimes as $0.80 \le \beta_w \le 0.974$ and the true
wavelength increases by a factor $1.22 = 0.974/0.80$. The observed
component spacing associated with the apparent helical wavelength would
increase according to equation (4) by a factor 6.6 from component $s$
to component $o$, or from, say, 0.4 to 2.6~mas consistent with the
observed component spacing.  The corresponding true wavelength,
$\lambda$, varies from 0.35~pc near the core to 0.43~pc at 8.4~pc
(2~mas observed) from the core where the jet radius, $R_{jt}$, is $\approx
0.15$~pc ($\approx 0.21$~mas), and $\lambda \sim 3 R_{jt}$.

A wave speed of $\beta_w = 0.974$ in the high frequency regime and flow
speed $\beta = 0.986$ implies a jet sound speed $a_{jt} = 0.3 c$ according
to equation (3).  A wave speed of $\beta_w = 0.80$ in the low frequency
regime implies $a_{ex}/a_{jt} = 0.35$ and $a_{ex} = 0.1 c$ from equation (1a).
These derived parameters are very similar to those used for the
``warm'' jet case shown in the previous sections. The corresponding
dispersion relation solution panel in Figure 1 ($a_{ex} = 0.081 c$)
clearly shows that $\lambda \sim 3 R_{jt}$ is in the high frequency regime
as assumed.  The wavelength and wave speed inferred from the
observations imply a frequency $\omega R_{jt}/a_{ex} \sim 60$ ($\omega
R_{jt}/v_{jt} \sim 6$) at component $o$ ($z \sim 8.4$~pc).

On a strictly isothermal jet, the dispersion relation solution implies
a jet expansion by a factor $\approx 40$ beginning at $z \sim 0.21$~pc
to achieve the observed acceleration in wave speed.  This position is
inside component $s$, has $\omega R_{jt}/a_{ex} \sim 1.5$ and implies an
initial excitation at about the resonant frequency.  In this case the
speed of component $s$ may more closely represent the wave speed at
resonance.  Unfortunately the estimated wave speed at resonance given
by equation (2a) is insufficiently accurate for our purposes and, in
general, a numerically found value must be used. For the present
parameters the wave speed at resonance is slightly less than in the low
frequency regime and thus $a_{ex} \gtrsim 0.1$~c would be implied.  Note
that the inferred jet parameters allow for an apparent faster motion of
components closer to the core, if the helical path could be
followed/existed inside component $s$.

The amplitude growth curve for the high frequency helical wave on a
``warm'' jet (Figure 3 top right panel) gives a good indication of
expected wave growth for a helical wave beyond the resonant location
independent of the exact location or mechanism for the initial
perturbation.  Here we have a resonant location $z_0 \equiv 0.21$~pc
where $R_0 = 3.7 \times 10^{-3}$~pc.  Saturation from an initial
amplitude $A_0/R_0 << 0.07$ can easily be achieved by $z = z_0 +
3.7$~pc ($\lesssim 1$~mas) where the saturation amplitude is $A/R_{jt} <
0.025$.

\vspace{-0.8cm}
\section{Summary and Conclusion}

The present work has shown that helical structure can be sensitive
to jet and external medium conditions.  This sensitivity manifests
itself in predictable changes in wave speed, wavelength and growth rate
along an expanding jet.  The basic behavior of a non-disruptive
``linear'' helical twist is summarized below:

\begin{enumerate}

\vspace{-0.2 cm} 
\item  The wave speed varies, typically between limits set by the value of $\gamma a_{ex}/a_{jt}$ at low and $a_{jt}$ at high frequencies, relative to a resonant frequency, $\omega^* R_{jt}/a_{ex} \gtrsim 1.5$, at which the growth rate is largest.

\vspace{-0.2 cm}
\item  The accompanying long and short wavelength regimes scale relative to the resonant wavelength $\lambda^* \propto \{ [\gamma (M_{jt}^2-1)^{1/2}]/[\gamma (M_{jt}^2 -1)^{1/2} + (M_{ex}^2 - 1)^{1/2}] \} M_{ex} R_{jt}$.

\vspace{-0.2 cm}
\item The spatial growth length scales $\propto \gamma M_{jt}$.  A helical twist is damped along an expanding jet if $\gamma M_{jt}$ is too large.

\vspace{-0.2 cm}
\item Typically, a fixed frequency helical wave propagates from the low frequency to high frequency regimes by virtue of jet expansion.

\vspace{-0.2 cm}
\item  Given equal excitation of multiple frequencies at a position $z_0$ the
strongest helical pattern at $z$ will be at a frequency $\omega
> \omega^*$ and wavelength $\lambda < \lambda^*$.

\end{enumerate}

Modeling the detailed flow and pressure structure accompanying a
helical twist shows the highest pressures to be near the surface of the
jet.  In general the jet fluid does not follow the helical path of the
high pressure ridge.  The correspondence between jet surface
displacement, and pressure and velocity structure is summarized below:

\begin{enumerate}

\vspace{-0.2 cm}
\item A given pressure fluctuation induces a larger velocity fluctuation for higher jet sound speed. 

\vspace{-0.2 cm}
\item  Surface displacements accompanying similar levels of pressure
fluctuation are larger for lower jet sound speeds but remain relatively
small on relativistic flows.

\vspace{-0.2 cm}
\item   Maintaining a saturation amplitude implies a decrease in the surface displacement as $\omega > \omega^*$.

\vspace{-0.2 cm}
\item At saturation amplitudes the jet center is at higher speed and lower pressure than the jet surface.

\vspace{-0.2 cm}
\item  Axial velocity fluctuations are largest near the jet surface. Transverse velocity fluctuations are largest in the jet interior. 

\vspace{-0.2 cm}
\item  The maximum helical flow pitch angle is less than the relativistic Mach angle and always less than the helical pressure ridge pitch angle.

\end{enumerate}

The present work shows how projection and light travel time effects
profoundly influence the appearance of helical twisting for jets at
small viewing angles.  This extreme sensitivity should allow the
determination of jet and external medium properties, albeit it is
essential to have proper motion information in order to constrain the
observed jet's speed and orientation.  The brightest radio emission is
expected to be generated in the region of highest pressure so
observations will trace the path of the high pressure ridge.  On an
expanding relativistic jet the theoretically predicted patterns are
like those that would be expected if the jet intensity were enhanced
along a helical path on a conical surface.  The brightest regions are
where the helical path approaches closer to the viewing angle. The
brightness is enhanced by foreshortening and by Doppler boosting,
albeit by less than would be assumed from the pitch angle of the high
pressure ridge.  The observed appearance of a helical twist is
summarized below:

\begin{enumerate}

\vspace{-0.2 cm}
\item Short wavelength twists in the high frequency regime travel at
nearly the jet speed.  On superluminal jets their apparent wavelength
will be longer than the intrinsic wavelength.

\vspace{-0.2 cm}
\item  For a fixed frequency helical wave the apparent wavelength
change is amplified by $(1 - \beta_w~cos~\theta_0)^{-1}$ relative to
the intrinsic wavelength change $\propto \beta_w$.  As a result
$\lambda^{obs}$ can appear $\propto R_{jt}$ as a jet expands if $a_{ex} \ll a_{jt}$.

\vspace{-0.2 cm}
\item  Multiple frequencies can produce $\lambda^{obs} \propto R_{jt}$ on average and can produce very different apparent motions at comparable locations.

\end{enumerate}

The motion of components in the 3C\,120 jet inside 10~mas may serve as an
example of the predicted helical wave behavior on an expanding
relativistic jet.  At an assumed viewing angle of 9.6\arcdeg ~the
observed component motions would indicate an isothermal jet expansion
with jet sound speed $\approx 0.3$~c and sound speed in the medium
immediately outside the jet $\gtrsim 0.1$~c. Additional modeling work
is planned to constrain the viewing angle and evaluate the robustness
of these estimates.

It is anticipated that detailed modeling of observed helical structure
on jets with observed proper motions will provide strong constraints on
jet and external sound speeds, and on the perturbation frequencies that
arise from or near to the central engine.  Note that considerably
different wave speed and wavelength behavior is associated with, for
example, the present assumed isothermal as opposed to an adiabatic
expansion.  Thus, modeling results have direct implications for
particle energization/heating rates and indirectly for particle
acceleration rates.  The finding of say high jet sound speed implies
the lack of a ``cold'' baryonic jet component and may provide
constraints on the composition of jet material, i.e., electron-proton
dominated or electron-positron dominated as some polarization
observations can be taken to imply (Wardle et al.\ 1998; Celotti et
al.\ 1998).

\acknowledgements 
The author would like to thank Jean Eilek, Andrei Lobanov and Craig
Walker for motivating this study and providing illuminating
discussions.   The author would also like to thank Jose-Luis G\'omez
for helpful comments and for providing an image of the inner 3C\,120
jet.  P. Hardee acknowledges partial support from the National Science
Foundation through grant AST-9802955 to the University of Alabama.

\end{document}